\documentclass[journal]{IEEEtran}
\usepackage{cite}
\usepackage{graphicx}
\usepackage[final]{pdfpages}
\usepackage{amsmath}
\usepackage{mathdots}
\usepackage{amsfonts}
\usepackage{mathrsfs}      
\usepackage{amssymb}

\begin{document}
  \title{Gray Box Identification of State-Space Models Using Difference of Convex Programming}
  \author{Chengpu Yu, Lennart Ljung and Michel Verhaegen%
  \thanks{C. Yu and M. Verhaegen are with the Delft Center for Systems and Control, Delft University, Delft 2628CD, Netherlands (c.yu-4@tudelft.nl, m.verhaegen@tudelft.nl)}
  \thanks{L. Ljung is with the Division of Automatic Control, Department of Electrical Engineering, Link¨oping University, Sweden (ljung@isy.liu.se)}
  \thanks{This work is sponsored by the European Research Council, Advanced Grant Agreement No. 339681.}
  }
  \maketitle
\begin{abstract}
Gray-box identification is prevalent in modeling physical and networked systems. However, due to the non-convex nature of the gray-box identification problem, good initial parameter estimates are crucial for a successful application. In this paper, a new identification method is proposed by exploiting the low-rank and structured Hankel matrix of impulse response. This identification problem is recasted into a difference-of-convex programming problem, which is then solved by the sequential convex programming approach with the associated initialization obtained by nuclear-norm optimization. The presented method aims to achieve the maximum impulse-response fitting while not requiring additional (non-convex) conditions to secure non-singularity of the similarity transformation relating the given state-space matrices to the gray-box parameterized ones. This overcomes a persistent shortcoming in a number of recent contributions on this topic, and the new method can be applied for the structured state-space realization even if the involved system parameters are unidentifiable. {The method can be used both for directly estimating the gray-box parameters and for providing initial parameter estimates for further iterative search in a conventional gray-box identification setup.}
\end{abstract}

\section{Introduction}
Nowadays, the control and identification of structured state-space system model have attracted great attention in the control community. There are two main sources of structured state-space models: the modeling of practical physical systems \cite{dorf2011modern,ljung1998system,verhaegen2007filtering} and the description of networked systems \cite{bellman1970structural,van1998structural}. When modeling physical systems, the non-zero entries of the system matrices always have physical meanings such as masses, velocity, acceleration, and so on. Identification of the {physical} parameters can provide us a  better understanding of the inner physical mechanism of the investigated object. On the other hand, a network connected system often can be represented as a structured system with the structure straightforwardly determined by the interconnections among the involved subsystems. Identification of such kind of structured system models provides the foundation for the model-based network control.

In the literature, there are two kinds of methods to identify structured state-space models. \emph{One} is  {the traditional gray-box set-up}, to identify the parameterized state-space models directly from the input-output (IO) data using the prediction-error method \cite{ljung1998system},\cite{verhaegen2007filtering}. Since the involved identification problem is always a non-convex optimization problem, the conventional nonlinear optimization methods, such as regularized Gauss-Newton method {\cite[Section 10.2]{ljung1998system}}, and the gradient project method \cite[Chapter 7]{verhaegen2007filtering}, are sensitive to the initial {parameter estimate}. {This traditional setup thus requires reasonable knowledge of the parameters and structures to be identified. Since the gray box situation starts from some physical insight, this knowledge may be sufficient in some cases, but too demanding in other. Resorting to testing random initial parameters may not be feasible for large problems.}

\emph{The other} {approach to structured state-space models} is to {first estimate an unstructured, black-box model}
 using, e.g., subspace identification methods, followed by the recovery of the  {physical}  parameters embedded in the concerned structured model. Using the classical subspace identification methods, such as MOESP and N4SID \cite{ljung1998system,verhaegen2007filtering}, the system matrices in the first step can be consistently estimated under some mild conditions. The parameter recovery in the second step turns out to be a small-scale bilinear optimization problem.

To solve the bilinear optimization problem involved with the gray-box identification, an alternating minimization algorithm was developed in \cite{xie2002estimate} and a null-space based method was proposed in \cite{prot2012null}. In order to prevent the singular similarity transformation, a non-smooth optimization approach was presented in \cite{mercere2014identification}. Furthermore, in order to avoid estimating the similarity transformation, an $H_{\infty}$-norm-based identification algorithm was proposed in \cite{Vizer201634}. The above mentioned algorithms are sensitive to initial conditions. To cope with this problem, the bilinear optimization problem was reformulated into a sum-of-squares of polynomials which is then solved by the semi-definite programming method \cite{ljung2003initialization}; however, this technique is limited to solving small-scale problems having only a few unknown variables.

In this paper, a difference-of-convex programming (DCP) based method is developed for the identification of structured state-space models. This approach estimates the system parameters by the structured factorization of a block Hankel matrix of system impulse response, which is inspired by the Ho-Kalman decomposition method \cite{verhaegen2007filtering}. More explicitly, the proposed method boils down to solving a low-rank structured matrix factorization problem. In this paper, this non-convex optimization problem is transformed into a difference-of-convex programming (DCP) problem which is then tackled by the sequential convex programming technique \cite{boyd2008sequential}.

The advantages of the proposed method against many recently developed methods are as follows. Different from the identification method in \cite{ mercere2014identification}, the proposed algorithm framework avoids the non-singularity constraint on the similarity transformation and can be applied to the realization of non-identifiable gray-box models. Unlike the model-matching $H_{\infty}$ method \cite{Vizer201634} which requires to solve an infinite-dimensional optimization problem, the proposed identification method is finite-dimensional so that it is more computational amenable. Moreover, compared with other gradient-based or alternating minimization methods \cite{xie2002estimate,verhaegen2007filtering}, the proposed identification method performs  well in practice thanks to the high-quality initial conditions obtained by solving the convexified low-rank and structured matrix factorization problem.

The rest of the paper is organized as follows. Section \ref{sec1} formulates the identification problem of gray-box models. Section \ref{sec2} reviews the gray-box identification using the classical prediction-error method. Section \ref{sec3} gives an alternative way for the gray box identification, which is to identify the black-box model first, following the identification of system parameters by solving a bilinear optimization problem. Section \ref{sec4} provides a new method for solving the bilinear optimization problem. Section \ref{sec5} demonstrates the performance of the proposed identification method, followed by some conclusions in Section \ref{sec6}.

\section{Problem statement}\label{sec1}
We consider a parameterized state-space model as follows
\begin{equation}\label{sec1_equ1}
  \begin{split}
    &\dot x(t)=A(\theta)x(t)+B(\theta)u(t)\\
    &y(kT)=C(\theta)x(kT)+w(kT),
  \end{split}
\end{equation}
where $u(t)\in\mathbb R^{m},x(t)\in\mathbb R^{n}, y(t)\in\mathbb R^{p}$ and $w(k)\in\mathbb R^{p}$ are system input, state, output, and measurement noise, respectively; $\theta\in\mathbb R^{l}$ is the parameter vector; $t$ and $k$ represent continuous and discrete time indices, respectively; $T$ is the sampling period. \\

The parameter vector $\theta$ in \eqref{sec1_equ1} typically represents unknown values of physical coefficients. Here, we assume that the structured system matrices are affine with respect to $\theta$, i.e.
\[
  \begin{split}
    &A(\theta)=A_0+\sum_{i=1}^l A_i\theta_i\\
    &B(\theta)=B_0+\sum_{i=1}^l B_i\theta_i\\
    &C(\theta)=C_0+\sum_{i=1}^l C_i\theta_i,
  \end{split}
\]
where the coefficient matrices $A_i,B_i$ and $C_i$ are known. Besides the structures of the system matrices, the system order of \eqref{sec1_equ1} is known as a priori knowledge as well.\\

Denote the corresponding true continuous-time transfer function by:
\begin{equation}
  G(s,\theta)=C(\theta)\left(sI-A(\theta)\right)^{-1}B(\theta).
\end{equation}
Although state-evolution equation in \eqref{sec1_equ1} is continuous, we can only obtain sampled IO data in practice with sampling period $T$. Denoting the discrete-time system, obtained by the sampling period $T$ with the system input $u(t)$ being piecewise constant between the sampling instants $kT$, as
\begin{equation}\label{sec1_equ2}
  H(q,\theta)=C(\theta)\left(qI-A_T(\theta)\right)^{-1}B_T(\theta),
\end{equation}
where
\[A_T=e^{A(\theta)T},\quad B_T=\int_{\tau=0}^T e^{A(\theta)\tau}B(\theta)d\tau.\]

Given the sampled IO data $\{u(kT),y(kT)\}$ for $k=0,1,\cdots$ that are generated from model \eqref{sec1_equ1} for a certain value $\theta^*$, the concerned gray-box identification problem is to estimate the parameter vector $\theta^*$ from the measured IO data. \\

To address the concerned identification problem, the following assumptions are made throughout the paper.
 \begin{itemize}
    \item[A1] The system in \eqref{sec1_equ1} is minimal;
    \item[A2] The magnitudes of the imaginary parts of the eigenvalues of $A(\theta)$ are less than the Nyquist frequency $\frac{\pi}{T}$.
    \item[A3] The input sequence $u(kT)$ is persistently exciting of any finite order \cite[Chapter 13]{ljung1998system};
    \item[A4] The measurement noise $w(t)$ is uncorrelated with the system input $u(t)$.
 \end{itemize}
 Assumption A2 ensures that the corresponding discrete-time model of \eqref{sec1_equ1} is minimal when Assumption A1 is satisfied \cite{chen1994optimal}. Assumptions A3-A4 are standard assumptions for the consistent identification of the discrete-time system model $H(q,\theta)$.\\

\section{Gray-box approach}\label{sec2}
The estimation of the parameter vector $\theta^*$ using the sampled IO data $\{u(kT),y(kT)\}$ is typically a \emph{gray-box identification problem}. The traditional identification method is the prediction-error method \cite{ljung1998system} in which the predicted or simulated outputs $\hat y(kT|\theta)$ are computed using the discrete-time model $H(q,\theta)$ for any $\theta$. The corresponding prediction error criterion is written as
\begin{equation}\label{sec2_equ1}
\begin{split}
  \min_{\theta}&\quad\frac{1}{N}\sum_{k=0}^{N-1}\|y(kT)-\hat y(kT{|\theta})\|^2\\
  s.t.&\quad \hat x(kT+T)=A_T(\theta)\hat x(kT)+B_T(\theta)u(kT)\\
  &\quad \hat y(kT{|\theta})=C(\theta)\hat x(kT)+w(kT)\textrm{ for }k=0,1,\cdots,N-1.
\end{split}
\end{equation}
This general method has the best possible asymptotic accuracy, but the main drawback is that the optimization problem is (highly) non-convex and may have many local minima. The gradient-based optimization algorithms such as Gauss-Newton method  {\cite[Section 10.2]{ljung1998system}},  and gradient projection method \cite[Chapter 7]{verhaegen2007filtering} can be used to solve \eqref{sec2_equ1}. However, the performance mainly relies on the selection of initial {parameter estimate. The gray-box structure information may be sufficient to provide such initial estimates that are in the domain of attraction of the global minimum but otherwise one may have to resort to random initial parameters.} It is shown in \cite{ljung2003initialization} that the chances to reach the global minimum of \eqref{sec2_equ1} from random starting points {may be} very slim for  problems of realistic sizes.

\section{Black-box + Algebraic Approach}\label{sec3}
Besides the gray-box approach, there exist other routes to estimate the parameter vector $\theta^*$ from the sampled IO data. Even though the gray-box approach may end up in local minima, it is still possible to find the true system from data by a black-box approach. Subspace approaches like N4SID and MOESP \cite{ljung1998system,verhaegen2007filtering} can, under mild conditions, obtain the true discrete-time system $H(q,\theta^*)$ as the length of the IO data tends to infinity. That discrete-time system
can be easily transformed to continuous-time using the zero-order hold interpolation approach \cite{franklin1998digital}. As a result, the continuous-time transfer function $G(s,\theta^*)$ will be known, but in an unknown state-space basis:
\begin{equation}
  G(s,\theta^*)=C^*(sI-A^*)^{-1}B^*.
\end{equation}
The identification problem has now been transformed to an algebraic problem:\\
\emph{Given the values of $A^*,B^*,C^*$, determine the parameter vector $\theta$ satisfying}
\begin{equation}\label{sec2_equ2}
C^*(sI-A^*)^{-1}B^*=C(\theta)(sI-A(\theta))^{-1}B(\theta).
\end{equation}
{The estimate of $\theta$ obtained in this way can then be used as initial estimate in the minimization of (\ref{sec2_equ1}).}
This approach was discussed in \cite{xie2002estimate,ljung2003initialization,mercere2014identification}.\\

\section{Solving the Algebraic problem}\label{sec4}
To solve the algebraic problem in \eqref{sec2_equ2}, two routes are provided here: one is the similarity transformation of the state-space realization and the other is the low-rank and structured factorization of the block Hankel matrix of impulse response.

\subsection{Using Similarity Transformation}
Equation \eqref{sec2_equ2} means that there exists a similarity transformation $Q$ such that
\begin{equation}\label{sec3_equ2}
  QA^*=A(\theta)Q,\quad QB^*=B(\theta),\quad C^*=C(\theta)Q.
\end{equation}
From that we can form the criterion
\begin{equation}\label{sec3_equ1}
  \begin{split}
    V(Q,\theta)&=\|QA^*-A(\theta)Q\|_F^2+\|QB^*-B(\theta)\|_F^2\\
    &\quad +\|C^*-C(\theta)Q\|_F^2
  \end{split}
\end{equation}
 The optimization problem in \eqref{sec3_equ1} is a bilinear estimation problem and an alternating minimization method was proposed in \cite{xie2002estimate}. In \cite{ljung2003initialization}, the optimization problem in \eqref{sec3_equ1} was minimized by a convex sum-of-squares method in case $A(\theta),B(\theta),C(\theta)$ are affine in $\theta$; however, this method is limited to solving small-scale problems having only {rather} few
unknown variables. \\

 The equation group in \eqref{sec3_equ2} can be equivalently written in a vector form:
 \begin{equation}\label{equ5}
    \underbrace{\left[\begin{array}{c}
      (A^*)^T\otimes I-I\otimes A(\theta)\\
      (B^*)^T\otimes I\\
      I\otimes C(\theta)
    \end{array}\right]}_{M(\theta)}\textrm{vec}(Q)=\underbrace{\left[\begin{array}{c}
      0\\
      \textrm{vec}(B(\theta))\\
      \textrm{vec}(C^*)
    \end{array}\right]}_{N(\theta)}.
  \end{equation}
 To solve the above bilinear equation, a gradient projection method was given in \cite[Chapter 7.5.4]{verhaegen2007filtering}, a null-space-based optimization method was developed in \cite{prot2012null} and a difference-of-convex based method was proposed in \cite{yu2015identification}.\\

 Even if we can find a global optimal solution $(Q^{\star},\theta^{\star})$ by one of the above mentioned methods, it might not be meaningful for the identification purpose stated in Section \ref{sec1}. The reason for this is that the optimal solution $Q^{\star}$ might be singular and the obtained transfer function $C^{\star}(sI-A^{\star})^{-1}B^{\star}$ might not be equal to $C^{*}(sI-A^{*})^{-1}B^{*}$. In fact, equations \eqref{sec2_equ2} and \eqref{sec3_equ2} are equivalent if and only if $Q$ is nonsingular \cite{kailath1980linear}. To deal with this problem, a condition-number constraint on $Q$ was considered in \cite{mercere2014identification}, which turns out to be a non-smooth and highly non-convex optimization problem.\\

 Another way to deal with the possible mismatch between equations \eqref{sec2_equ2} and \eqref{sec3_equ2} is to minimize the model-matching criterion $\|G(s,\theta)-G(s,\theta^*)\|$ using either $H_2$ norm or $H_{\infty}$ norm, as suggested in \cite{ljung2003initialization}. The $H_{\infty}$-norm based model-matching method has been investigated in \cite{Vizer201634}. Compared with the minimization of \eqref{sec3_equ1}, the $H_{\infty}$-norm based model-matching method reduces the number of unknown variables but at the price of the introduction of a semi-infinite and non-smooth program \cite{Vizer201634}.\\

\subsection{Using the Hankel Matrix of Impulse Response }
In this section, aiming at dealing with the possible drawback of minimizing equation \eqref{sec3_equ1}, a new identification approach is developed by exploiting the structured and low-rank factorization of the block Hankel matrix of impulse response.\\

After obtaining a full-parameterized state-space realization $G(s,\theta^*)=C^*(sI-A^*)^{-1}B^*$, we can obtain the associated impulse response sequence denoted by
\[
M_i(\theta^*)=C(\theta^*)A^i(\theta^*)B(\theta^*)=C^*(A^*)^iB^*
\]
for $i=0,1,\cdots$. Let $H_{v,h}(\theta^*)$ be a block Hankel matrix constructed by Markov parameters
\begin{equation}\label{sec4_equ0}
  H_{v,h}(\theta^*)=\left[\begin{array}{cccc}
    M_0(\theta^*)&M_1(\theta^*)&\cdots&M_{h-1}(\theta^*)\\
    M_1(\theta^*)&M_2(\theta^*)&\cdots&M_h(\theta^*)\\
    \vdots&\vdots&\iddots&\vdots\\
    M_{v-1}(\theta^*)&M_v(\theta^*)&\cdots&M_{v+h-2}(\theta^*)
  \end{array}\right],
\end{equation}
where the subscripts $v,h$, satisfying $v,h\geq n$, denote the number of block rows and number of block columns, respectively. Given the block Hankel matrix $H_{v,h}(\theta^*)$, the concerned gray-box identification problem is formulated as
\begin{equation}\label{hankel_formulation}
\small\begin{split}
  \min_{\theta}&\quad \|H_{v,h}(\theta^*)-H_{v,h}(\theta)\|_F^2\\
  s.t.&\quad H_{v,h}(\theta)=\left[\begin{array}{ccc}
    C(\theta)B(\theta)&\cdots&C(\theta)A^{h-1}(\theta)B(\theta)\\
    \vdots&\iddots&\vdots\\
   C(\theta)A^{v-1}(\theta)B(\theta)&\cdots&C(\theta)A^{v+h-2}(\theta)B(\theta)
  \end{array}\right].
\end{split}
\end{equation}

In the above equation, the block Hankel matrix $H_{v,h}(\theta)$ has a low-rank factorization as
\begin{equation}\label{low_rank_form}
\begin{split}
  H_{v,h}(\theta)&=\underbrace{\left[\begin{array}{c}
    C(\theta)\\
    C(\theta)A(\theta)\\
    \vdots\\
    C(\theta)A^{v-1}(\theta)
  \end{array}\right]}_{\mathcal O_v(\theta)}\\
  &\times\underbrace{\left[\begin{array}{cccc}
    B(\theta)&A(\theta)B(\theta)&\cdots&A^{h-1}(\theta)B(\theta)
  \end{array}\right]}_{\mathcal C_h(\theta)},
\end{split}
\end{equation}
where $\mathcal O_v(\theta)$ and $\mathcal C_h(\theta)$ denote the extended observability and controllability matrix, respectively.\\

Denote $Y=H_{v,h}(\theta^*)$. By exploiting the shift properties embedded in extended observability and controllability matrices, the optimization problem \eqref{hankel_formulation} can be recasted into a low-rank structured matrix factorization problem:
\begin{equation}\label{shift_formulation}
\begin{split}
  &\min_{\theta,\mathcal O_v,\mathcal C_h, X}\quad \|Y-X\|_F^2\\
  &s.t.\quad X=\mathcal O_v\mathcal C_h\\
      &\quad\quad\mathcal O_v\left(1:p,:\right)=C(\theta)\\
      &\quad\quad\mathcal C_h\left(:,1:m\right)=B(\theta)\\
      &\quad\quad\mathcal O_v\left(1:(v-1)p,:\right)A(\theta)=\mathcal O_v\left(p+1:vp,:\right)\\
      &\quad\quad A(\theta)\mathcal C_h\left(:,1:(h-1)m\right)=\mathcal C_h\left(:,m+1:hm\right).
\end{split}
\end{equation}
In the above optimization problem, the first and the last two constraints in the above equation are bilinear. To solve this problem, the DCP-based identification framework \cite{yu2015identification} will be adopted, which contains the following three steps: (i) the bilinear optimization problem is transformed into a rank constrained optimization problem; (ii) the rank constrained problem is recasted into a DCP problem; (iii) the DCP problem is then solved using the sequential convex programming technique.\\

\emph{Step 1:}
 The first constraint, $X=\mathcal O_v\mathcal C_h$, in \eqref{shift_formulation} can be equivalently written as a rank constraint.
\newtheorem{lm_rank}{Lemma}
\begin{lm_rank}\label{bilinear_low_equ}
  \cite{reinier2016ecc} The bilinear equation $X=\mathcal O_v\mathcal C_h$ is equivalent to the rank constraint
  \begin{equation}
    \textrm{rank}\left[\begin{array}{cc}
      X&\mathcal O_v\\
      \mathcal C_h& I_n
    \end{array}\right]=n.
  \end{equation}
\end{lm_rank}

 The equivalent rank constraints for the last two constraints of \eqref{shift_formulation} are derived below.  To simplify the notation, we denote $\bar{\mathcal O}_v=\mathcal O_v(1:(v-1)p,:),\underline{\mathcal O}_v=\mathcal O_v(p+1:vp,:)$ and $\bar{\mathcal C}_h=\mathcal C_h(:,1:(h-1)m),\underline{\mathcal C}_h=\mathcal C_h(:,m+1:hm)$.
The last two constraints can be represented as
\begin{equation}\label{vect_prod}
 \begin{split}
   &\bar{\mathcal O}_vA_0+\sum_{i=1}^q \left(\bar{\mathcal O}_v\theta_i\right) A_i=\underline{\mathcal O}_v\\
   &A_0\bar{\mathcal C}_h+\sum_{i=1}^q A_i\left(\bar{\mathcal C}_h\theta_i\right) =\underline{\mathcal C}_h.
 \end{split}
\end{equation}
An equivalent form of the combination of the fourth and fifth constraints is given in the following lemma.
\newtheorem{lm_rank_one}[lm_rank]{Lemma}
\begin{lm_rank_one}\label{rank_one_lemma}
  The constraint equation \eqref{vect_prod} is equivalent to
  \begin{equation}\label{rank_one_equ}
    \begin{split}
      &\bar{\mathcal O}_vA_0+\sum_{i=1}^q \Gamma_i A_i=\underline{\mathcal O}_v\\
      &A_0\bar{\mathcal C}_h+\sum_{i=1}^q A_i\Upsilon_i =\underline{\mathcal C}_h\\
      &\textrm{rank}\left[\begin{array}{cccc}
      1&\theta_1&\cdots&\theta_q\\
        \textrm{vec}(\bar{\mathcal O}_v)&\textrm{vec}(\Gamma_1)&\cdots&\textrm{vec}(\Gamma_q)\\
        \textrm{vec}(\bar{\mathcal C}_h)&\textrm{vec}(\Upsilon_1)&\cdots&\textrm{vec}(\Upsilon_q)
      \end{array}\right]=1.
    \end{split}
  \end{equation}
\end{lm_rank_one}
\begin{proof}
  To prove the lemma, it is sufficient to prove that the variables in \eqref{vect_prod} and those in \eqref{rank_one_equ} are one-to-one mapping. On the one hand, the variables in \eqref{rank_one_equ} can be uniquely determined from those in \eqref{vect_prod} by assigning $\Gamma_i=\bar{\mathcal O}_v\theta_i$ and $\Upsilon_i=\bar{\mathcal C}_h\theta_i$. On the other hand, the variables in \eqref{vect_prod} can be uniquely determined by the SVD decomposition of the matrix on the left-hand side of the following equation
  \[
  \small\left[\begin{array}{cccc}
        1&\theta_1&\cdots&\theta_q\\
        \textrm{vec}(\bar{\mathcal O}_v)&\textrm{vec}(\Gamma_1)&\cdots&\textrm{vec}(\Gamma_q)\\
        \textrm{vec}(\bar{\mathcal C}_h)&\textrm{vec}(\Upsilon_1)&\cdots&\textrm{vec}(\Upsilon_q)
      \end{array}\right]=\left[\begin{array}{c}
      1\\
        \textrm{vec}(\bar{\mathcal O}_v)\\
        \textrm{vec}(\bar{\mathcal C}_h)
      \end{array}\right]\left[\begin{array}{cc}
        1&\theta^T
      \end{array}\right].
  \]
\end{proof}

By Lemmas \ref{bilinear_low_equ} and \ref{rank_one_lemma}, the bilinear optimization problem in \eqref{shift_formulation} can be equivalently formulated as a rank-constrained optimization problem as follows:
\begin{equation}\label{whole_rank_opt}
  \begin{split}
   & \min_{\theta,\mathcal O_v,\mathcal C_h, X,\Gamma,\Upsilon}\quad \|Y-X\|_F^2\\
  &s.t.\;\textrm{rank}\left[\begin{array}{cc}
      X&\mathcal O_v\\
      \mathcal C_h& I_n
    \end{array}\right]=n\\
      &\quad \mathcal O_v\left(1:p,:\right)=C(\theta)\\
      &\quad \mathcal C_h\left(:,1:m\right)=B(\theta)\\
      &\bar{\mathcal O}_vA_0+\sum_{i=1}^q \Gamma_i A_i=\underline{\mathcal O}_v\\
      &\bar{\mathcal O}_v=\mathcal O_v(1:(v-1)p,:),\;\;\underline{\mathcal O}_v=\mathcal O_v(p+1:vp,:)\\
      &A_0\bar{\mathcal C}_h+\sum_{i=1}^q A_i\Upsilon_i =\underline{\mathcal C}_h\\
      &\bar{\mathcal C}_h=\mathcal C_h(:,1:(h-1)m),\;\;\underline{\mathcal C}_h=\mathcal C_h(:,m+1:hm)\\
      &\textrm{rank}\left[\begin{array}{cccc}
      1&\theta_1&\cdots&\theta_q\\
        \textrm{vec}(\bar{\mathcal O}_v)&\textrm{vec}(\Gamma_1)&\cdots&\textrm{vec}(\Gamma_q)\\
        \textrm{vec}(\bar{\mathcal C}_h)&\textrm{vec}(\Upsilon_1)&\cdots&\textrm{vec}(\Upsilon_q)
      \end{array}\right]=1.
  \end{split}
\end{equation}

The above optimization contains two rank constraints. To deal with the above rank constrained optimization, we shall further formulate it as a difference of convex optimization problem.\\

\emph{Step 2:}
For notational simplicity, we denote
\[
T=\left[\begin{array}{cccc}
        1&\theta_1&\cdots&\theta_q\\
        \textrm{vec}(\bar{\mathcal O}_v)&\textrm{vec}(\Gamma_1)&\cdots&\textrm{vec}(\Gamma_q)\\
        \textrm{vec}(\bar{\mathcal C}_h)&\textrm{vec}(\Upsilon_1)&\cdots&\textrm{vec}(\Upsilon_q)
      \end{array}\right].
\]
Let $\sigma_i\left(T\right)$ be the $i$-th largest singular value of $T$ for $i=1,\cdots,q+1$. Define
\[
f_{\kappa}\left(T\right)=\sum_{i=1}^{\kappa} \sigma_i\left(T\right).
\]
It is remarked that $f_{\kappa}(\cdot)$ is a Ky Fan $\kappa$-norm, which is a convex function \cite{qi1996extreme}.

Inspired by the truncated nuclear norm method in \cite{hu2013fast,yu2015identification}, the rank constraint $\textrm{rank}\left(T\right)=1$ can be replaced by
\begin{equation}
  f_{q+1}\left(T\right)-f_{1}\left(T\right)=0.
\end{equation}
The above equation means that all the singular values except the largest one of $T$ are zero. Since $f_{q+1}(T)=\|T\|_*$, the above equation can be represented as
\[
\|T\|_*-f_1\left(T\right)=0.
\]

Using the above strategy, instead of directly solving the rank constrained optimization problem in \eqref{whole_rank_opt}, we try to solve the following regularized optimization problem:
\begin{equation}\label{regu_opt}
   \small\begin{split}
    &\min_{\theta,\mathcal O_v,\mathcal C_h, X,\Gamma,\Upsilon}\quad \|Y-X\|_F^2+\lambda_1\left(\|\Gamma\|_*-f_n(\Gamma)\right)\\
    &\quad\quad\quad\quad+\lambda_2\left(\|T\|_*-f_1(T)\right)\\
 & s.t.\quad\Gamma=\left[\begin{array}{cc}
      X&\mathcal O_v\\
      \mathcal C_h& I_n
    \end{array}\right]\\
      &\quad \mathcal O_v\left(1:p,:\right)=C(\theta)\\
      &\quad \mathcal C_h\left(:,1:m\right)=B(\theta)\\
      &\bar{\mathcal O}_vA_0+\sum_{i=1}^q \Gamma_i A_i=\underline{\mathcal O}_v\\
      &\bar{\mathcal O}_v=\mathcal O_v(1:(v-1)p,:),\;\;\underline{\mathcal O}_v=\mathcal O_v(p+1:vp,:)\\
      &A_0\bar{\mathcal C}_h+\sum_{i=1}^q A_i\Upsilon_i =\underline{\mathcal C}_h\\
      &\bar{\mathcal C}_h=\mathcal C_h(:,1:(h-1)m),\;\;\underline{\mathcal C}_h=\mathcal C_h(:,m+1:hm)\\
      &T=\left[\begin{array}{cccc}
      1&\theta_1&\cdots&\theta_q\\
        \textrm{vec}(\bar{\mathcal O}_v)&\textrm{vec}(\Gamma_1)&\cdots&\textrm{vec}(\Gamma_q)\\
        \textrm{vec}(\bar{\mathcal C}_h)&\textrm{vec}(\Upsilon_1)&\cdots&\textrm{vec}(\Upsilon_q)
      \end{array}\right],
  \end{split}
\end{equation}
where $\lambda_1,\lambda_2$ are regularization parameters. It is remarked that all the constraints in \eqref{regu_opt} are linear functions with respect to the unknown variables and the objective function is a difference-of-convex function. Although the formulations \eqref{whole_rank_opt} and \eqref{regu_opt}  may not be strictly equivalent, they have the same global optimum.\\

\emph{Step 3:} We shall develop a sequential convex programming method to solve the DC optimization problem in \eqref{regu_opt}. In order to develop a sequential convex programming method, it is essential to linearize the concave terms in the objective function of \eqref{regu_opt}. Let $\Gamma^j$ be the estimate of $\Gamma$ at the $j$-th iteration and its SVD decomposition be given as
\begin{equation}\label{svd_gamma}
  \Gamma^j=\left[\begin{array}{cc}
    U_1^j& U_2^j
  \end{array}\right]\left[\begin{array}{cc}
    S_1^j\\
    & S_2^j
  \end{array}\right]\left[\begin{array}{c}
    V_1^{j,T}\\
    V_2^{j,T}
  \end{array}\right],
\end{equation}
where $U_1^j$ and $V_1^j$ are respectively the left and right singular vectors corresponding to the largest $n$ singular values. It can be established that \cite{qi1996extreme}
\begin{equation}
  U_1^{j}V_1^{j,T}\in\partial f_n\left(\Gamma^j\right).
\end{equation}
Then, the linearization of $f_n(\Gamma)$ at the point $\Gamma=\Gamma^j$ is
\begin{equation}\label{lin_gamma}
  f_n(\Gamma)\approx f_n(\Gamma^j)+\textrm{tr}\left(U_1^{j,T}\left(\Gamma-\Gamma^j\right)V_1^j\right).
\end{equation}

Let  $T^j$ be the estimate of $T$ at the $j$-th iteration and its SVD decomposition be given as
\begin{equation}\label{svd_t}
 T^j=\left[\begin{array}{cc}
    L_1^j& L_2^j
  \end{array}\right]\left[\begin{array}{cc}
    \Sigma_1^j\\
    & \Sigma_2^j
  \end{array}\right]\left[\begin{array}{c}
    R_1^{j,T}\\
    R_2^{j,T}
  \end{array}\right],
\end{equation}
where $L_1^j$ and $R_1^j$ are respectively the left and right singular vectors corresponding to the largest singular value. Then, the linearization of $f_1(T)$ at the point $T=T^j$ is
\begin{equation}\label{lin_t}
  f_1(T)\approx f_1(T^j)+\textrm{tr}\left(L_1^{j,T}\left(T-T^j\right)R_1^j\right).
\end{equation}

Based on the linearizations in \eqref{lin_gamma} and \eqref{lin_t}, the convex optimization problem to be solved at the $(j+1)$-th iteration is as follows:

\begin{equation}\label{seq_conv}
   \small\begin{split}
    &\min_{\theta,\mathcal O_v,\mathcal C_h, X,\Gamma,\Upsilon}\quad \|Y-X\|_F^2+\lambda_1\left(\|\Gamma\|_*-\textrm{tr}\left(U_1^{j,T}\Gamma V_1^j\right)\right)\\
    &\quad\quad\quad+\lambda_2\left(\|T\|_*-\textrm{tr}\left(L_1^{j,T}TR_1^j\right)\right)+\rho\left(\|\Gamma-\Gamma^j\|_F^2+\|T-T^j\|_F^2\right)\\
  &s.t.\;\Gamma=\left[\begin{array}{cc}
      X&\mathcal O_v\\
      \mathcal C_h& I_n
    \end{array}\right]\\
      &\quad \mathcal O_v\left(1:p,:\right)=C(\theta)\\
      &\quad \mathcal C_h\left(:,1:m\right)=B(\theta)\\
      &\bar{\mathcal O}_vA_0+\sum_{i=1}^q \Gamma_i A_i=\underline{\mathcal O}_v\\
      &\bar{\mathcal O}_v=\mathcal O_v(1:(v-1)p,:),\;\;\underline{\mathcal O}_v=\mathcal O_v(p+1:vp,:)\\
      &A_0\bar{\mathcal C}_h+\sum_{i=1}^q A_i\Upsilon_i =\underline{\mathcal C}_h\\
      &\bar{\mathcal C}_h=\mathcal C_h(:,1:(h-1)m),\;\;\underline{\mathcal C}_h=\mathcal C_h(:,m+1:hm)\\
      &T=\left[\begin{array}{cccc}
       1&\theta_1&\cdots&\theta_q\\
        \textrm{vec}(\bar{\mathcal O}_v)&\textrm{vec}(\Gamma_1)&\cdots&\textrm{vec}(\Gamma_q)\\
        \textrm{vec}(\bar{\mathcal C}_h)&\textrm{vec}(\Upsilon_1)&\cdots&\textrm{vec}(\Upsilon_q)
      \end{array}\right],
  \end{split}
\end{equation}
where $\rho$ is a very small positive regularization parameter and the proximal term $\rho\left(\|\Gamma-\Gamma^j\|_F^2+\|T-T^j\|_F^2\right)$ is added to ensure the convergence of the sequential convex programming approach, as shown in Lemma \ref{lm_conver}.\\

To ease the reference, the above sequential convex programming procedure is summarized in Algorithm 1.
\begin{table}[ht]
\centering 
\begin{tabular}{l}
\hline
\textbf{Algorithm 1} Sequential convex programming method for \eqref{regu_opt}\\
\hline
1) Set $\Gamma^0=0$ and $T^0=0$.\\
2) Repeat\\
    $\quad$ 2-1): Compute respectively the left and right singular vectors of  $\Gamma^j$ \\
    $\quad\quad\quad$ and $T^j$ as shown in  \eqref{svd_gamma} and \eqref{svd_t}. \\
    $\quad$ 2-2): Obtain the estimates $\Gamma^{j+1}$ and $T^{j+1}$ by solving \eqref{seq_conv}.\\
3) until $\frac{\|\theta^{j+1}-\theta^{j}\|_2}{\|\theta^j\|_2}\leq \varepsilon$ with $\varepsilon$ a small value.\\
\hline
\end{tabular}
\end{table}

By applying the iterative optimization method in Algorithm 1, the convergence is guaranteed.
\newtheorem{lm_converg}[lm_rank]{Lemma}
\begin{lm_converg}\label{lm_conver}
  For the difference-of-convex optimization problem in \eqref{regu_opt}, by implementing Algorithm 1, the obtained estimate $\theta^j$ satisfies that $\lim_{j\rightarrow\infty} \left(\theta^{j+1}-\theta^j\right)=0$. Any accumulation point of $\{\theta^j\}$ is a stationary point of \eqref{regu_opt}.
\end{lm_converg}
The above lemma can be proven by the results of Theorems 1-2 in \cite{lu2016nonconvex}.\\

Since the difference-of-convex optimization problem in \eqref{regu_opt} is still non-convex, the performance of the provided sequential convex programming procedure relies on the initial conditions. However, by setting $T^0=0$ and $\Gamma^0=0$, we can find that the optimization problem in \eqref{regu_opt} is a nuclear-norm relaxation of the rank-constrained optimization problem in \eqref{whole_rank_opt}. Due to the fact that the nuclear norm is the convex envelope of the low-rank constraint on the unit spectral norm ball \cite{recht2010guaranteed}, the associated nuclear-norm optimization usually yields a good candidate for the starting point of the sequential convex programming procedure.

\section{Numerical Simulations}\label{sec5}
In this section, two simulation examples will be carried out to demonstrate the performance of the proposed method - Algorithm 1. For comparison purposes, the prediction-error method (PEM) {\cite{ljung1998system},} \cite{ljung2003initialization} and the difference-of-convex programming  (DCP) method \cite{yu2015identification} are simulated. The implementation details of these three methods are given below.
\begin{enumerate}
  \item Algorithm 1 is simulated by empirically setting the regularization parameters in \eqref{seq_conv} to $\lambda_1=10^{-4}$, $\lambda_2=10^{-5}$ and $\rho=10^{-10}$. The tolerance of relative error is set to $\epsilon=10^{-4}$.
  \item PEM is simulated by firstly configuring the structure object using the Matlab command {\texttt{idgrey}, see \cite{sitb83}} and then implementing the identification method using the Matlab command {\texttt{pem}}. The initial conditions are randomly generated following the standard Gaussian distribution.
  \item DCP is simulated by setting the regularization parameter $\lambda$ in equation (17) of \cite{yu2015identification} to $\lambda=10^{-4}$. The tolerance of relative error is set to $10^{-4}$.
\end{enumerate}
In the simulations, the maximum number of iterations for the these three methods is set to 100.

\subsection{Randomly generated structures}
The first simulation example is conducted following the way in \cite{ljung2003initialization}. The state-space model of \eqref{sec1_equ1} is randomly generated by the Matlab command \texttt{rss}, and the system parameters to be estimated are randomly picked from the generated models. Since the system model is randomly generated, it is difficult to find a unified sampling period for all systems. Therefore, when simulating DCP and Algorithm 1, the system matrices $A^*,B^*$ and $C^*$ in \eqref{sec2_equ2} are assumed to be known.

To ensure the identifiability of the system parameters, the number of unknown parameters cannot be larger than $(p+m)n$; however, system parameters less than $(p+m)n$ may not always be identifiable \cite{ljung1998system,verhaegen2007filtering}. Therefore, we use the \emph{impulse-response fitting} to measure the identification performance.
{In the simulation, we choose the system order $n=5$ and the input/output dimension $m=p=1$. For each fixed number of free parameters, we carry out 100 Monte-Carlo trials by randomly generating the system model and randomly picking a fixed number of free parameters. The success rate is obtained by counting the number of successful trials using the criterion $\textrm{IRF}^r\leq 10^{-3}$.}
 Denote by $\theta^r$ the estimate of $\theta$ at the $r$-th Monte-Carlo trial. The impulse-response fitting (IRF) of the $r$-th Monte-Carlo trial is defined as
\begin{equation}
 \textrm{IRF}^r=\frac{\sum_{i=0}^{v+h-2}\|C(\theta^r)A^i(\theta^r)B(\theta^r)-C^*(A^*)^iB^*\|_F}{\sum_{i=0}^{v+h-2}\|C^*(A^*)^iB^*\|_F},
\end{equation}
where the dimension parameters $v$ and $h$ are defined in \eqref{sec4_equ0}.

\begin{figure}
  \centering
  \includegraphics[scale=0.5]{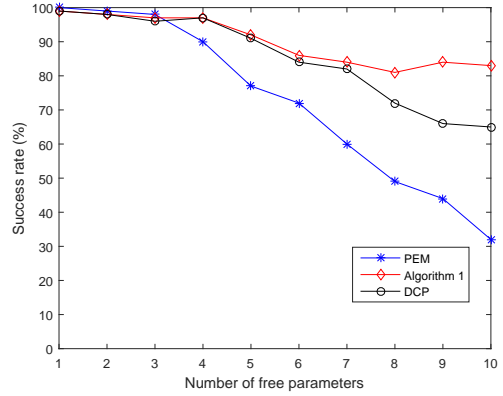}
  \caption{Example 1: identification performance of PEM, DCP and Algorithm 1.}\label{figure1}
\end{figure}

The identification performance of PEM, DCP and Algorithm 1 is shown in Fig. \ref{figure1} from which we can draw the following conclusions.
\begin{enumerate}
  \item When the number of parameters is larger than 3, DCP and Algorithm 1 perform much better than PEM. This is because DCP and Algorithm 1 can find good initializations by nuclear-norm regularized optimization. In contrast, when the number of parameters are less than or equal to 3, PEM has a slightly better performance than DCP and Algorithm 1. This might be relevant to the selection of the regularization parameters of DCP and Algorithm 1.
  \item When the number of parameters is larger than 6, the success rate of Algorithm 1 is higher than that of DCP up to 20\%. This might be caused by the fact that DCP does not consider the non-singularity constraint of the similarity transformation, while Algorithm 1 implicitly guarantees the non-singularity of the similarity transformation. However, when the number of parameters is less than or equal to 6, DCP and Algorithm 1 have similar performance. This might be because, when the number of free parameters becomes smaller, singular similarity transformations are less likely to occur.
\end{enumerate}
\subsection{Compartmental structures}
The second simulation example is to identify the compartmental structure having the following form \cite{bellman1970structural}
\begin{equation}
 \scriptsize\begin{split}
  &A(\theta)=\left[\begin{array}{ccccccc}
    -\theta_1&\theta_2\\
    \theta_1&-(\theta_2+\theta_3)&\ddots\\
    &\ddots&\ddots&\ddots\\
    &&\ddots&-(\theta_{2n-4}+\theta_{2n-3})&\theta_{2n-2}\\
    &&&\theta_{2n-3}&-\theta_{2n-2}
  \end{array}\right]\\
  &B=\left[\begin{array}{cccccc}
    0&0&\cdots&0&1
  \end{array}\right]^T,\quad \quad C=\left[\begin{array}{cccccc}
    0&0&\cdots&0&1
  \end{array}\right].
  \end{split}
\end{equation}
For each fixed system order, 100 Monte-Carlo trials are carried out by randomly generating the system parameters. The success rate is obtained by counting the number of successful trials using the criterion $\textrm{IRF}^r\leq 10^{-3}$.

\begin{figure}
  \centering
  \includegraphics[scale=0.5]{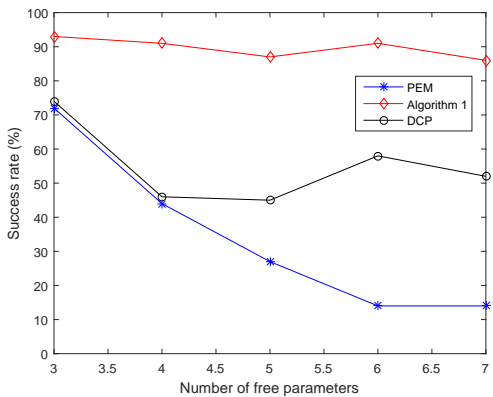}
  \caption{Example 2: identification performance of PEM, DCP and Algorithm 1.}\label{figure2}
\end{figure}

Fig. \ref{figure2} shows the identification performance of three investigated methods in terms of IRF. It can be found that the success rates of Algorithm 1 stay around 90\% for different system orders, which demonstrates the better performance of Algorithm 1 with relation to DCP and PEM. This can be explained by that Algorithm 1 can obtain good initializations and guarantee the non-singularity of the similarity transformation.

\section{Conclusions}\label{sec6}
In this paper, we have proposed a new gray-box identification method by exploiting the low-rank and structured factorization of the Hankel matrix of impulse response. This method uses the system impulse-response fitting as the objective function while avoiding the explicit non-singularity constraint on similarity transformation; thus, it can be applied to the state-space realizations of non-identifiable gray-box models. Compared with the classical prediction-error method initialized at random parameter values, the proposed method can yield better performance since it can find a good initialization by nuclear-norm based optimization.

Although the proposed identification algorithm demonstrates good performance in terms of system impulse-response fitting, its computational complexity is higher than the classical prediction-error method. Thus, investigation will be made in our future work on improving the computational efficiency of the proposed identification method.

 \bibliographystyle{ieeetr}
 \bibliography{ref_file}
\end{document}